\documentclass[conference]{IEEEtran}
\IEEEoverridecommandlockouts
\usepackage{cite}
\usepackage{amsmath,amssymb,amsfonts}
\usepackage{algorithmic}
\usepackage{graphicx}
\usepackage{textcomp}
\usepackage{xcolor}
\usepackage{url}
\usepackage{subfig}
\usepackage{subfloat}
\def\BibTeX{{\rm B\kern-.05em{\sc i\kern-.025em b}\kern-.08em
    T\kern-.1667em\lower.7ex\hbox{E}\kern-.125emX}}
\begin{document}

\title{From AI-assisted to AI-native Industrial Control: Architecture, Agent Design, and Challenges\\
\thanks{This work was supported by the Major Key Project of PCL, China [PCL2025A13]. Corresponding author: Shengkai Liu (liushengkai@cuhk.edu.cn)}
}

\author{\IEEEauthorblockN{1\textsuperscript{st} Lisi Liu}
\IEEEauthorblockA{\textit{Department of New Networks} \\
\textit{Pengcheng Laboratory}\\
Shenzhen, China \\
liuls@pcl.ac.cn}
\and
\IEEEauthorblockN{2\textsuperscript{nd} Bin Wang}
\IEEEauthorblockA{\textit{Department of New Networks} \\
\textit{Pengcheng Laboratory}\\
Shenzhen, China \\
wangb02@pcl.ac.cn}
\and
\IEEEauthorblockN{3\textsuperscript{rd} Quan Tang}
\IEEEauthorblockA{\textit{Department of New Networks} \\
\textit{Pengcheng Laboratory}\\
Shenzhen, China \\
bebdong@gmail.com}
\and
\IEEEauthorblockN{4\textsuperscript{th} Chuan Pang}
\IEEEauthorblockA{\textit{Department of New Networks} \\
\textit{Pengcheng Laboratory}\\
Shenzhen, China \\
pangch@pcl.ac.cn}
\and
\IEEEauthorblockN{5\textsuperscript{th} Feng Gao}
\IEEEauthorblockA{\textit{Department of New Networks} \\
\textit{Pengcheng Laboratory}\\
Shenzhen, China \\
gaof01@pcl.ac.cn}
\and
\IEEEauthorblockN{6\textsuperscript{th} Shengkai Liu}
\IEEEauthorblockA{\textit{Department of New Networks} \\
\textit{Pengcheng Laboratory}\\
Shenzhen, China \\
liushengkai@cuhk.edu.cn}
}

\maketitle

\begin{abstract}
The integration of AI and industrial control is rapidly transforming industrial automation. However, current approaches remain predominantly AI-assisted, where AI functions are incorporated as auxiliary components without fundamentally altering the underlying architecture of industrial control systems. Consequently, the potential of AI to enable higher levels of autonomy, adaptability, and continuous evolution remains underutilized. This paper proposes an AI-native industrial control paradigm, which reconceptualizes industrial control systems by positioning AI as the intrinsic core rather than an external enhancement. The proposed paradigm is built upon an AI-native infrastructure and a collection of collaborative agentic AIs, each encapsulating a set of advanced skills. Among them, the industrial control agent plays a central role in orchestrating the system lifecycle. Through this paradigm, industrial control shifts from a closed, capability-oriented, human-driven, and passive engineering process to an open, task-oriented, intent-driven, and autonomous solution delivery mechanism. The paper further analyzes the key challenges of the AI-native paradigm from both AI and industrial control perspectives, including issues related to AI model reliability, formal verification, real-time assurance, and heterogeneous resource orchestration. Overall, this work outlines a forward-looking vision for industrial control in the AI era and identifies critical research directions toward autonomous, evolvable, and adaptive control systems in complex and dynamic industrial environments.
\end{abstract}


\section{Introduction}
Artificial Intelligence (AI) is a significant driver of the ongoing industrial transformation, reshaping the manufacturing lifecycle from design and engineering to production and maintenance. Within the production stage, the integration of AI into the industrial control system (ICS), such as programmable logic controller (PLC) and computer numerical controller (CNC), is accelerating, primarily manifesting in two modalities:

(a) AI-in-the-loop, wherein specialized AI modules (e.g., visual inspection) are embedded within control logic to couple probabilistic inference with deterministic execution \cite{mas,deng}; 

(b) AI-assisted operation, which leverages generative AIs as copilots to automatically generate or optimize control programs from natural language specifications, significantly reducing development time and human error, and employs AI for installation, diagnostics, predictive maintenance, and other operational guidance, improving efficiency and reliability \cite{liu,vv1,vv3}.

Despite these advancements, the prevailing integration of AI in ICS remains ``AI-assisted'' rather than ``AI-native''. In this paradigm, the traditional ICS constitutes the core, with AI features functioning as peripheral enhancements. That is, the AI-assisted paradigm represents a mode of augmentation over reinvention, layering new capabilities onto a legacy foundation without prompting a fundamental redesign of its underlying architecture and operating model. Such a compromising development strategy is pragmatic but limited in autonomy, evolvability, and adaptability. 

In the contemporary context of mass customization and volatile supply chains, traditional ICS's rote execution capability is no longer sufficient \cite{seitz,nag}. The control layer must be autonomous, evolving, dual-mode adaptable: it should execute top-down directives with determinism while responding to bottom-up, emergent events from the physical world \cite{bah}. Therefore, we argue that an evolutionary shift from the AI-assisted to an AI-native ICS is necessary. Rather than an AI-enhanced retrofit, it should be a system redesigned from first principles, in which AI and control are co-designed as intrinsic capabilities, so that they are deeply integrated from the outset.

In this context, this paper explores an AI-native ICS paradigm. Based on a brief reviewing of traditional ICS and AI-assisted ICS paradigm in Section~\ref{sec:review}, Section~\ref{sec:overview} introduces the proposed AI-native ICS paradigm, including architecture, key components, and demonstration. Section~\ref{sec:challenges} discusses the key research challenges from both AI and ICS perspectives, and Section~\ref{sec:conclusion} concludes the paper.

\section{A brief review of traditional ICS and AI-assisted ICS}
\label{sec:review}

\begin{figure}[tb]
\centerline{\includegraphics[width=0.45\textwidth]{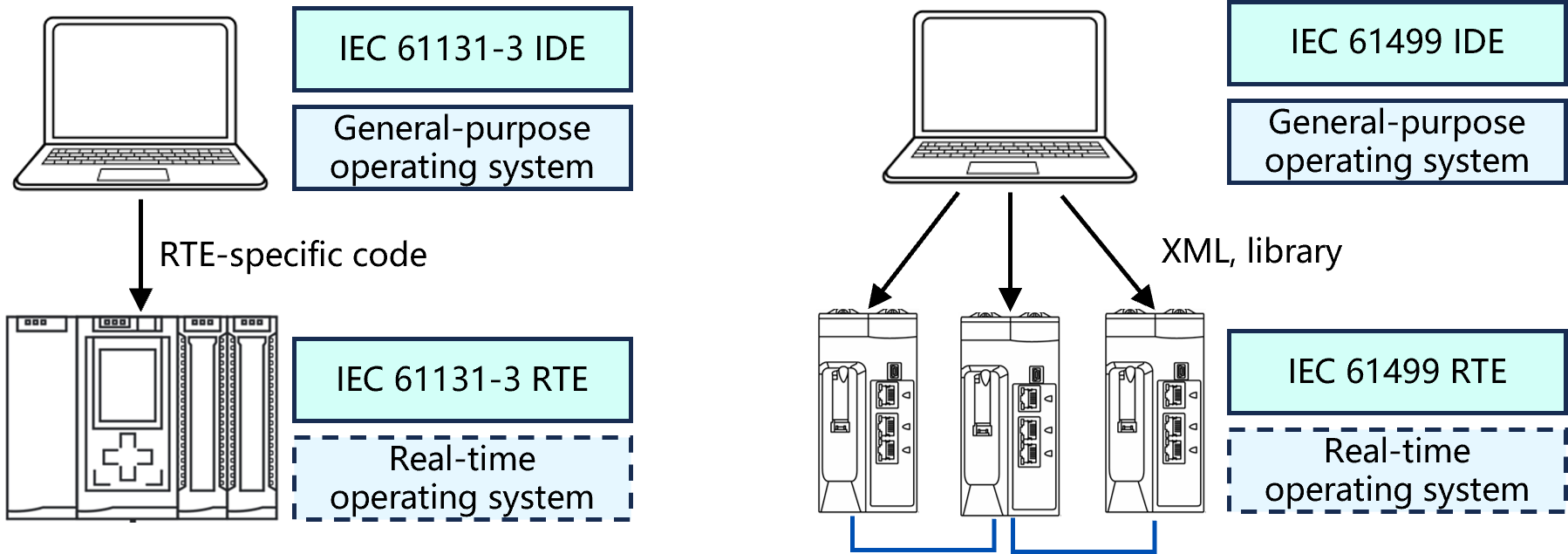}}
\caption{Architecture and execution mechanism of traditional PLC}
\label{tradition}
\end{figure}

This paper examines PLC as a representative archetype to review traditional ICS. Architecturally, a PLC consists of two tightly coupled components: the integrated development environment (IDE) and the runtime environment (RTE), namely, \textit{PLC=IDE+RTE}, as illustrated in Fig.~\ref{tradition}.

The PLC IDE is a vendor-specific engineering suite, typically hosted on a general-purpose platform such as a Windows PC, providing functions for programming, configuration, and diagnostics. The PLC RTE executes PLC programs. It is usually built upon a real-time execution substrate, including bare-metal environments, proprietary RTOSs, or general-purpose operating systems with real-time extensions. Commercial PLC RTEs are typically implemented as vertically integrated hardware-software systems designed to guarantee determinism, reliability, and safety.

The execution mechanism of PLCs mainly follows two IEC standards. Most deployed PLCs are based on IEC 61131-3, which adopts a synchronous scan-cycle execution model \cite{sehr}. In contrast, IEC 61499 introduces an event-driven execution model for reconfigurable and distributed systems \cite{lyu}.

The closed architecture of traditional PLCs has formed vendor-specific product families with divergent capabilities and limited interoperability \cite{sehr}. Consequently, it establishes a human-driven and passive operating model that remains dominant in ICS. This workflow generally consists of four phases:

\begin{itemize}
\item Selection. Engineers analyze automation requirements, including I/O scale, performance, connectivity, and hardware configuration, to select an appropriate PLC configuration. Practical decisions are usually determined by both technical suitability and long-term economic considerations.
\item Installation. Engineers install PLC modules, power supplies, and I/O devices, and complete electrical wiring according to engineering schematics.
\item Programming and virtual commissioning. Engineers develop control logic in vendor-specific IDEs based on IEC 61131-3 or IEC 61499. Offline simulation is commonly performed to detect logical errors before deployment.
\item On-site commissioning and running. After deployment to the PLC RTE, engineers tune parameters such as timers, counters, and PID loops to achieve desired performance. The PLC then executes a largely fixed automation task during operation.
\end{itemize}

In this operating model, any modification to the automation task passively triggers only the necessary engineering phases, such as reprogramming, hardware reconfiguration, or even reselection of the PLC platform.

In summary, traditional ICS exhibits two characteristics. First, it is built upon a closed architecture with fixed functionality, leading to vendor lock-in, poor interoperability, and limited third-party integration. Second, it adopts a passive operating model in which engineering follows a predefined lifecycle optimized for stability rather than autonomous adaptation. Together, these characteristics establish the vertically integrated ecosystem that dominates industrial automation today.

Consequently, traditional PLCs provide generalized, capability-oriented execution substrates with predefined performance envelopes, while end-users should develop and commission task-specific control programs by themselves. In other words, traditional ICS does not deliver task-oriented automation solutions directly, but instead provides a capability-oriented infrastructure upon which customized control logic must be engineered and validated.


The AI-assisted ICS paradigm, although an advancement, remains fundamentally rooted in this traditional architecture. AI functions are mainly introduced into the engineering toolchain to assist engineers during selection, programming, and commissioning, or integrated into control logic through vendor-specific interfaces. Therefore, AI primarily acts as a decision-support or augmentation mechanism rather than the system core. As a result, AI-assisted ICS still inherits the limitations of traditional ICS and cannot fully realize the autonomy, adaptability, and continuous evolution enabled by AI.





\section{Overview of the AI-native ICS paradigm}
\label{sec:overview}

This paper introduces an AI-native ICS paradigm, a reconceptualization of industrial automation from first principles. Instead of augmenting systems with AI features, this paradigm fields a system that is intrinsically AI-centered, autonomous, evolutionary, and adaptive. 

\subsection{Architecture of the AI-native ICS}
\label{sec:architecture}

\begin{figure}[tb]
\centerline{\includegraphics[width=0.4\textwidth]{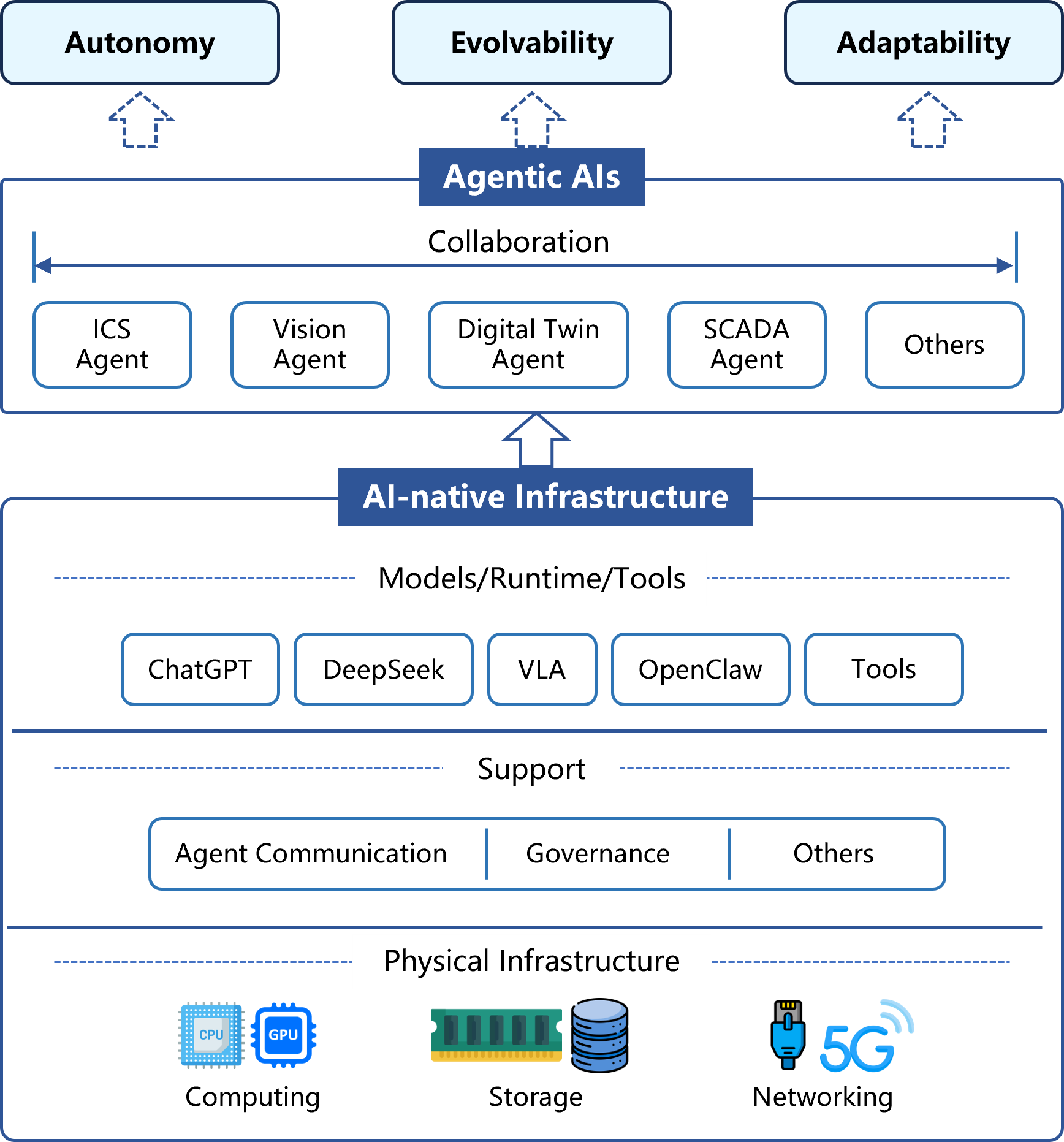}}
\caption{Conceptual architecture of the AI-native ICS}
\label{native}
\end{figure}

As illustrated in Fig.~\ref{native}, the conceptual architecture of this AI-native paradigm is composed of the AI-native infrastructure and a collection of collaborative agentic AIs. Diverging from the established cloud-native infrastructure \cite{cloud}, the AI-native infrastructure is architected on the premise that AI models and agents are the primary executive entities and that non-determinism is an inherent, manageable system characteristic, not an anomaly. From bottom to top, the AI-native infrastructure comprises: 
\begin{itemize}
    \item Physical Infrastructure, which includes heterogeneous computing, storage, and networking resources; 
    \item Support layer, which provides essential functions for agentic execution, such as inter-agent communication and resource scheduling; and 
    \item Models, Runtimes, and Tools layer, which encompasses a spectrum of AI models, specialized agent runtimes, and utilities for processes like fine-tuning and training.
\end{itemize}

This infrastructure provides the substrate for the execution and dynamic collaboration of multiple specialized agentic AIs, such as the ICS Agent, Vision Agent, and Digital Twin Agent. The agentic AI embodies three foundational characteristics: autonomy (self-governance), anagenesis (evolutionary advancement), and actionability (the capability to execute decisions). Through the collaboration of these agentic AIs, the ICS performs a high level of autonomy, evolvability, and adaptability.

\subsection{Key Skills of the ICS Agent and the Operating Model of the AI-native ICS}
\label{sec:skills}

The ICS agent is the entry point for the AI-native ICS. Rather than encapsulating traditional ICS as an agentic AI, it actively orchestrates the entire ICS lifecycle through a versatile suite of advanced skills. To illustrate, the key skills include (Fig.~\ref{plc}):

\begin{figure}[tb]
\centerline{\includegraphics[width=0.45\textwidth]{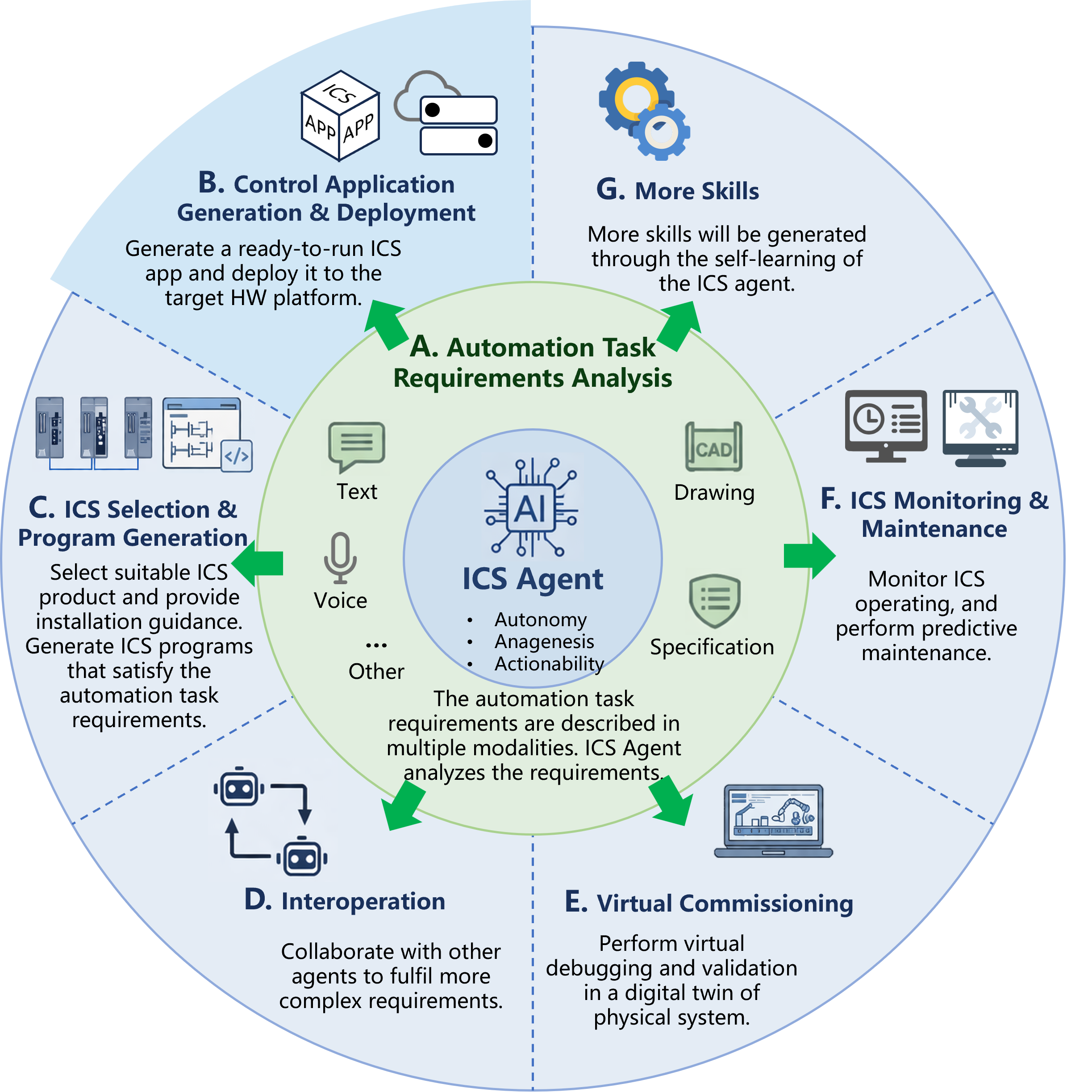}}
\caption{Significant skills of the ICS agent}
\label{plc}
\end{figure}

\textit{Skill A. Automation task requirements analysis}

This skill replaces the human-driven, passive process of translating requirements into rigid engineering specifications. Instead, the ICS agent directly interprets automation tasks from multimodal sources, such as textual descriptions, engineering drawings, or formal specifications. Through deep semantic understanding, it transforms this heterogeneous data into a unified, machine-executable internal representation. This skill establishes the basis for all subsequent agent skills. 

\textit{Skill B. On-demand control application generation and deployment}

As mentioned in Section~\ref{sec:review}, traditional ICS provides a capability-oriented solution based on the separation between IDE and RTE. This skill removes this separation by delivering a direct task-oriented solution. 

Rather than selecting a predefined RTE and coding corresponding control logic (programs), the ICS agent directly generates a task-specific control application tailored to the given automation requirements. When requirements evolve, it can automatically modify or regenerate the control application to maintain alignment with updated requirements. Besides, this skill takes responsibility for deploying the control application.

Overall, this skill aims to generate control applications and schedule computing and networking resources to execute control applications, rather than generating control programs targeting specific RTEs. As a result, it transforms ICS from a static, toolchain-based engineering solution into a dynamic, intent-driven, and self-adaptive task-oriented solution.

\textit{Skill C. Legacy system compatibility}

While the AI-native paradigm is forward-looking, it provides a pragmatic bridge to existing industrial ecosystems. To ensure backward compatibility, the ICS agent can select an optimal legacy RTE configuration based on requirements and generate installation guidance (or commands). Then, it can generate a tailored control program (e.g., in IEC 61131-3 or IEC 61499) for that specific RTE configuration. This capability allows the AI-native paradigm to subsume the functions of the AI-assisted model when necessary, offering a non-disruptive migration path.

\textit{Skill D. Agents interoperation}

The ICS agent operates within a broader ecosystem of specialized agents (e.g., MES agent, SCADA agent) to handle complex automation tasks. For instance, to execute the installation commands mentioned in Skill C, the ICS agent can issue formalized commands to an industrial robot agent. This robot agent then physically performs the installation and wiring, automating a task traditionally reliant on manual human labor. Similarly, the ICS agent can proactively respond to requests initiated by other agentic AIs, enabling dynamic and cooperative automation across the entire industrial landscape.

\textit{Skill E. Virtual commissioning}

By interacting with a digital twin agent, the ICS agent performs virtual commissioning before physical deployment. It validates, debugs, and optimizes the control logic within a high-fidelity virtual environment, drastically reducing the time, cost, and risks associated with on-site commissioning and ensuring a more robust initial deployment.

\textit{Skill F. Proactive monitoring and maintenance}

Post-deployment, the ICS agent continuously monitors the live system. It detects anomalies, predicts failures, and can even trigger autonomous maintenance procedures (e.g., by coordinating with other agents). This closed-loop operation ensures high availability and peak performance throughout the system's lifecycle.

\textit{Skill G. Anagenesis: continual skill and performance evolution}

The ICS agent's capabilities are not constant; its core architectural principle is anagenesis, or evolutionary self-improvement. By continuously learning from operational data, simulation outcomes, and interactions across the agent ecosystem, the ICS agent refines its existing skills and can develop entirely new ones. This inherent anagenesis ensures the AI-native ICS is a living system that evolves over time, a stark contrast to the fixed-function nature of traditional ICS.

\subsection{Demonstration of the AI-native ICS}
\label{sec:demo}

\begin{figure}[tb]
\centerline{\includegraphics[width=0.45\textwidth]{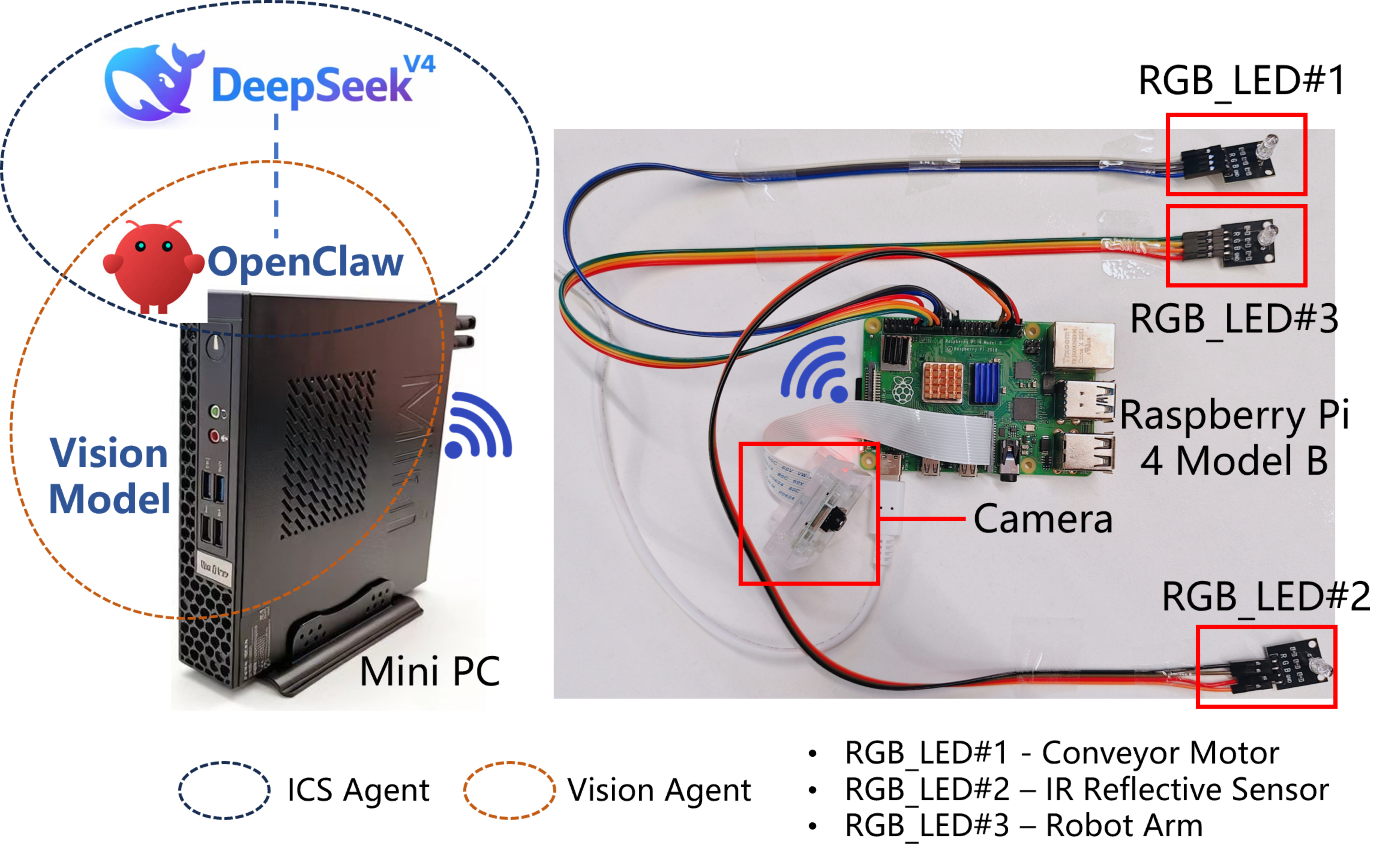}}
\caption{Sorting scenario under the AI-native ICS paradigm}
\label{example}
\end{figure}

\begin{figure}[tb]
    \centering
    \subfloat[Finite state machine diagram of Case \#1]{
    \label{fig:case1}
    \includegraphics[width=0.45\textwidth]{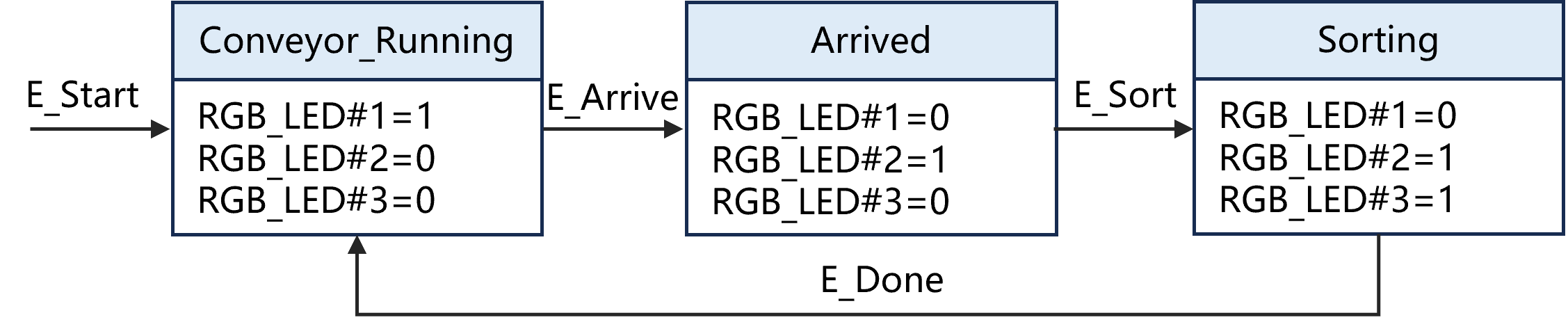}
    }
    \\
    \subfloat[Finite state machine diagram of Case \#2]{
    \label{fig:case2}
    \includegraphics[width=0.45\textwidth]{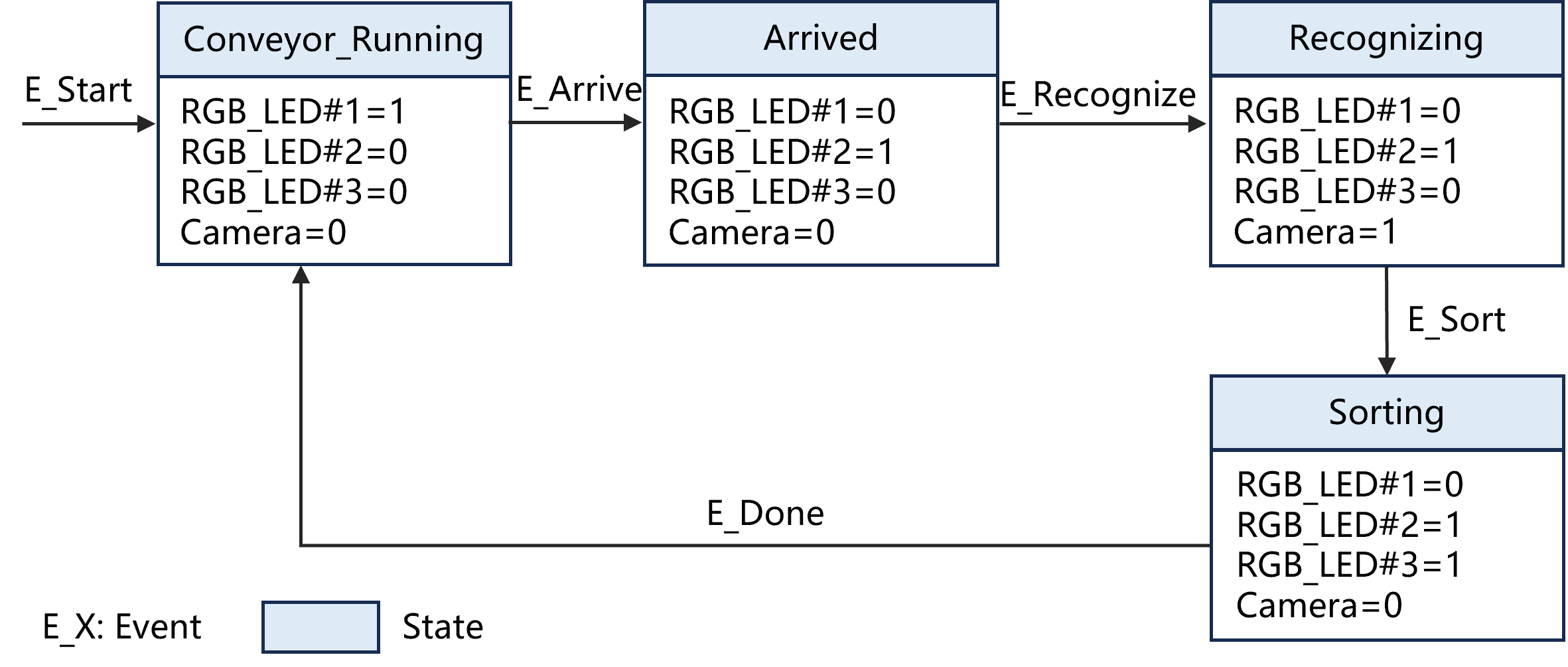}
    }
    \caption{Finite state machine diagrams of two cases' control logic}
    \label{fig:case}
\end{figure}

Figure.~\ref{example} illustrates a sorting scenario under the proposed AI-native ICS paradigm. Two representative cases are considered. In Case \#1, only a single type of product is transported on the conveyor. Once the product arrives at a designated position, the robot arm transfers it to a predefined storage location, as illustrated in Fig.~\ref{fig:case1}. In Case \#2, multiple types of products are transported on the conveyor. When a product reaches the designated position, a camera together with a vision model identifies the product type, after which the robot arm moves the product to the corresponding storage location, as illustrated in Fig.~\ref{fig:case2}.



In this demonstration, three RGB LEDs are employed to emulate the conveyor motor, the infrared reflective sensor, and the robot arm, respectively. Two agentic AI components are involved, namely the ICS agent and the Vision agent. Both agents are supported by OpenClaw deployed on a mini PC equipped with an Intel i5-7300H CPU and 8 GB DDR4 memory. Specifically, the ICS agent is currently implemented using DeepSeek V4, while the Vision agent is implemented using a lightweight vision model deployed locally on the mini PC. All RGB LEDs and the camera are connected to a Raspberry Pi 4 Model B, which resides in the same VLAN as the mini PC. In this scenario, the control applications generated by the ICS agent are deployed and executed on the Raspberry Pi 4 Model B.

For both cases, the control logic and hardware configuration are provided to the ICS agent in textual form. For example, the prompt for Case \#1 is expressed as follows:

\textit{``Generate an event-driven sorting control application. There are three RGB LEDs. LED\#1 represents the CONVEYOR\_RUNNING state, and its RGB pins are configured as red=17, green=22, and blue=27; LED\#2 ... Deploy and execute the generated application on the Raspberry Pi with IP address XXX ...''}

For Case \#2, additional perception-related requirements are incorporated into the prompt, for example:

\textit{``In this case, various types of products exist on the conveyor. The color of LED\#2 represents different product types ... In the RECOGNIZING state, the camera captures an image of LED\#2, and the corresponding color is identified by the vision model through interface XXX ... Modify the event-driven sorting control application accordingly ...''}

Table~\ref{tab:result} summarizes the experimental validation results and compares the proposed AI-native workflow with the conventional PLC-based engineering. For both cases, all operation phases were successfully completed without manual intervention after the task prompt was submitted.

\begin{table*}[t]
\caption{Experimental Validation and Comparison Between the Conventional PLC and the Proposed AI-Native ICS}
\label{tab:result}
\centering
\renewcommand{\arraystretch}{1.25}
\setlength{\tabcolsep}{5pt}
\begin{tabular}{|p{0.17\textwidth}|p{0.25\textwidth}|p{0.39\textwidth}|p{0.1\textwidth}|}
\hline
\textbf{Lifecycle Phase} &
\textbf{Conventional PLC} &
\textbf{Proposed AI-Native ICS} &
\textbf{Result} \\
\hline

Requirement analysis &
Automation requirements are manually analyzed by engineers. &
The ICS agent automatically interprets textual task requirements. &
Successfully completed \\
\hline

Control logic generation &
PLC programs are manually developed or generated with AI-assisted programming tools. &
The ICS agent automatically generates a task-specific event-driven control application. &
Successfully completed \\
\hline

Deployment &
The application is manually compiled, downloaded, and started on the PLC. &
The ICS agent automatically deploys and launches the generated application on the Raspberry Pi. &
Successfully completed \\
\hline

Execution &
The control logic is executed by a predefined PLC runtime. &
The generated control application is executed directly in the target hardware environment. &
Successfully completed \\
\hline

Requirement evolution &
Requirement changes require PLC reprogramming, redeployment, and recommissioning. &
The ICS agent regenerates, redeploys, and executes the control application according to the updated prompt. &
Successfully completed \\
\hline

\end{tabular}
\end{table*}

From this demonstration, moreover, it can be seen that rather than tasking AI with direct, real-time control - a proposition fraught with risk - the AI-native paradigm elevates AI to the role of an autonomous control architect. The ICS agent autonomously translates automation tasks into deterministic, verifiable control applications. Actuators remain governed by deterministic logic, ensuring operational safety and predictability. The significant advancement, therefore, is not in changing how the final actuator is controlled, but in fundamentally revolutionizing what it is told to do and how. Furthermore, driven by the collaboration of multiple agentic AIs, this new paradigm achieves unprecedented adaptivity in volatile industrial environments.


\section{Challenges of AI-native ICS paradigm}
\label{sec:challenges}

While the proposed AI-native paradigm unlocks potential capabilities in autonomy, evolvability, and adaptability, it introduces a new set of challenges. Due to the deep integration of AI and ICS, these challenges are examined from both the AI and ICS perspectives.  

\subsection{Challenges From the AI Perspective}
\label{sec:ai-issues}

AI-native ICS relies on advanced AI models, including LLMs (e.g., ChatGPT and DeepSeek), vision-language-action (VLA) models, and other foundation models. Despite their strong capabilities, these models still suffer from limitations such as hallucination and limited context reasoning constraints, which may lead to unreliable decisions in long-horizon or complex tasks. In addition, agentic AI runtimes (e.g., OpenClaw) introduce concerns regarding safety, privacy, and security \cite{mur,ran}. In ICS, such issues may directly propagate into control failures or even physical accidents. Therefore, ensuring the verifiability, predictability, and bounded behavior of agentic AI systems is a basic challenge for AI-native ICS. Potential solutions include constraining AI-generated actions, introducing formal or semi-formal verifications, and others. 

Moreover, the realization of AI-native ICS depends on AI-native infrastructure, which acts as the execution and coordination substrate for agentic AIs. Unlike cloud-native infrastructure, AI-native infrastructure is still at an early stage and lacks unified architectures and standards \cite{thota}. The layered architecture shown in Fig.~\ref{native} should therefore be regarded only as a conceptual reference. Open issues include the standard construction of agentic AIs, standardized communication mechanisms \cite{guo}, toolchains for training and inference, and resource scheduling across heterogeneous infrastructures. Consequently, both the architecture and implementation of AI-native infrastructure remain challenges for AI-native ICS.

\subsection{Challenges From the ICS Perspective}
\label{sec:ics-issues}

AI-native ICS envisions an autonomous and adaptive control ecosystem in which control logic is continuously generated and deployed according to changing requirements. However, traditional ICS is designed as a closed and vendor-specific ecosystem emphasizing stability, predefined functionality, and tightly controlled execution environments. This mismatch motivates Skill B (on-demand control application generation and deployment) in the ICS agent, while also introducing several challenges.

First, ensuring the correctness and real-time reliability of AI-generated control applications is a major issue. Unlike conventional IT software, control applications must satisfy strict requirements on determinism, latency, execution time, and functional correctness \cite{lei}. A promising direction is to integrate formal modelling and verification mechanisms into the generation process \cite{lilli}. For example, generated control logic can be mapped to timed automata, hybrid automata, or dependency models to verify timing constraints and reachability properties.


Second, deploying and executing generated control applications across heterogeneous infrastructures is another challenge. Unlike traditional ICS, where control logic is statically bound to a specific controller, the AI-native ICS enables more flexible deployment patterns. In AI-native ICS, control applications may be decomposed, distributed, or dynamically migrated across multiple platforms according to workloads and resource conditions. Although virtualization technologies provide a basis for decoupling control logic from hardware \cite{mak,jin,lyu2}, practical deployment remains difficult because control applications are often tightly coupled with CPU architectures, memory hierarchies, specialized accelerators (e.g., GPUs). Moreover, many control applications impose strict real-time constraints, making online migration and dynamic reallocation of resources highly non-trivial.

\subsection{Summary}
\label{sec:summary}
In summary, the challenges of the AI-native ICS paradigm stem from the fundamental tension between the probabilistic, evolving nature of AI and the deterministic, safety-critical requirements of ICS. From the AI perspective, issues such as hallucination, limited context reasoning, and the lack of verifiable guarantees directly affect the reliability of control-related decisions. From the ICS perspective, the AI-driven transition from a capability-oriented to a task-oriented solution introduces significant challenges for control-logic verification and infrastructure-resource orchestration. These challenges highlight that AI-native ICS is not merely an incremental extension of traditional systems, but a paradigm that integrates AI and ICS deeply. Addressing these challenges necessitates the integration of formal verification, real-time resource governance, and standardized AI-native infrastructure, thereby establishing a trustworthy foundation for the realization of autonomous, evolvable, and adaptive ICS.

\section{Conclusion}
\label{sec:conclusion}

This paper introduces an AI-native ICS paradigm, which reconceptualizes industrial automation from first principles by tightly integrating AI and ICS. Unlike the prevailing AI-assisted paradigm, where AI functions are externally grafted onto legacy architectures, the proposed paradigm treats AI as the intrinsic core of ICS, enabling autonomy, evolvability, and adaptability throughout the lifecycle. 

By analyzing traditional PLC-based ICS, this paper reveals its inherent limitations that impede the deep integration of AI and ICS, including closed, capability-oriented architecture solution, and human-driven, passive operating model. To address these limitations, the proposed AI-native paradigm introduces an agent-based architecture supported by AI-native infrastructure. In particular, the ICS agent is designed with a set of key skills, among which the on-demand control application generation and deployment capability fundamentally breaks the conventional IDE-RTE separation. This enables the direct synthesis of task-specific control applications from high-level automation intent, transforming ICS from a closed, capability-oriented solution to an open, task-oriented solution.

Furthermore, this paper analyzes the challenges of AI-native ICS from both AI and ICS perspectives. These challenges arise from the deep coupling between AI and ICS, including the need for verifiable and bounded AI behavior, as well as the difficulties in ensuring real-time determinism and orchestrating heterogeneous resources in dynamic execution environments. Therefore, future research should focus on formalizing control logic semantics, developing AI-native infrastructure, and establishing a trustworthy mechanism for verification and resource orchestration.








\end{document}